\documentclass[10pt,twocolumn]{IEEEtran}

\usepackage{float}
\usepackage{cite}
\usepackage{soul}
\usepackage{stackengine}
\usepackage{xcolor}

\usepackage{graphicx}
\usepackage{epstopdf}

\usepackage{subfig}
\begin{document}

\title{Phase sensitive parametric interactions in a $Ga_{0.5}In_{0.5}P$
photonic crystal waveguide}

\author{A.~Willinger\thanks{A. Willinger and G. Eisenstein are with the Department
of Electrical Engineering, Technion, Haifa 32000, Israel. e-mail: amnonjw@tx.technion.ac.il.}, A.~Martin\thanks{A. Martin is with Laboratoire de Photonique et Nanostructures, CNRS-UPR20, Marcoussis, France.}, S.~Combri\'{e}\thanks{S. Combrie and A. De Rossi are with Thales Research and Technology, Route D\'{e}partementale 128, 91767, Palaiseau, France.}, A.~De~Rossi, G.~Eisenstein,~\IEEEmembership{Fellow,~IEEE}}

\maketitle

\begin{abstract}
We report phase sensitive amplification in a $1.5mm$ long, $Ga_{0.5}In_{0.5}P$
dispersion engineered photonic crystal waveguide which has a flattened dispersion profile. A signal degenerate
configuration with pulsed pumps whose total peak power is only $0.5W$
yields a phase sensitive extinction ratio of $10dB$.\end{abstract}

%%\ociscodes{(050.5298) Photonic crystals; (060.4510) Optical communications; (190.4410) Nonlinear optics, parametric processes.}

\section{Introduction}
\IEEEPARstart{T}{he} concept of a phase sensitive parametric amplifier (PSA) was formulated
several decades ago \cite{Caves1982PhysRevD,Yuen1992ol}.
The parametric interaction in a PSA involves three fields with a well-defined
mutual phase relationship. The gain experienced by one of those fields
is maximized when it is in-phase with the other two fields; this field
is de-amplified when its phase is changed by $\pi/2$ radians. The
difference between maximum and minimum gain is defined as the phase-sensitive
extinction ratio (PSER). The noise accompanying the
amplification process is squeezed in one quadrature and the noise
figure can, in principle, reach a value of unity ($0dB$) \cite{Mu1992josab}.
Every PSA requires a preparatory stage of some kind \cite{Imajuku2000ElecLet,Tang2005opex,Tang2008opex,Neo2013opex}
where the mutual phases of the fields are properly arranged.

The fundamental concept of a PSA gained significant attention in recent
years as systems using phase encoded signals started to dominate fiber
optic communications. Indeed, PSAs were used in several experiments
where low noise amplification and phase noise regeneration were demonstrated
\cite{Tang2008opex,Slavik2010natpho,Tong2011natpho,Richter2012ofc}.
The use of a PSA in quantum optics where phase correlated spontaneous
photons are used for quantum information processing has also been
suggested \cite{Levandovsky1999ol}.

Most PSA demonstrations made use of optical fibers. Nonlinear interactions
in fibers are well understood; they can carry high optical powers
needed for the parametric interactions and they are naturally integrated
in fiber communication systems. The disadvantage of optical fibers
is their low nonlinearity which requires long interaction lengths
and hence special operating conditions have to be applied to prevent
Stimulated Brillouin and Raman scattering.

An attractive alternative to fiber based PSAs are compact waveguide
devices with large nonlinearities stemming from the materials used
and the small cross sections. A signal-degenerate PSA was demonstrated
in a $65mm$ long chalcogenide ridge waveguide \cite{Neo2013opex}
with $7ps$ pulses and a total pump peak powers of $7.3W$,
yielding a PSER of $10dB$. A similar device was used in the pump-degenerate
configuration \cite{Zhang2014josab} and achieved a higher PSER of
$18dB$ with a $6W$ peak pulse pump power. A $40mm$ long silicon waveguide
with a reverse-biased PIN junction \cite{DaRos2014opex} reached PSER
values between $10dB$ and $20dB$ with total peak pump powers of $0.25-0.63W$.
The PIN junction served to sweep out free carriers generated by two-photon
absorption (TPA) thereby reducing losses and increasing the nonlinear
efficiency. A recent report of a $20mm$ long SiGe ridge waveguide
\cite{Ettabib2015cleo} demonstrated a PSER of $28.6dB$ for
 total pump powers of only $0.14W$.

Since every parametric process is governed by phase matching,
dispersion control is a key feature. Photonic crystal waveguides (PCWs),
in particular dispersion engineered PCWs \cite{Monat2010opex,Colman2012opex}
are therefore ideal media for any parametric amplifier. A PSA based
on a pump degenerate configuration in a short, $196\mu m$, silicon
dispersion engineered PCW was demonstrated in \cite{Zhang2014ol}.
TPA limited operation even for such a short length so that a peak
pulse pump power of $2.3W$ was needed to reach a PSER of $10dB$.
Refractive index changes due to TPA induced free carriers were also
clearly observed.

In this paper, we report a PSA comprising a $1.5mm$ long $Ga_{0.5}In_{0.5}P$
dispersion engineered PCW. The advantage of $Ga_{0.5}In_{0.5}P$
is its large bandgap which prevents TPA at $1550nm$ and hence requires lower pump powers than similar silicon waveguides. Indeed, such a waveguide availed the first
phase insensitive parametric amplifier which provided an $11dB$ gain
in a $1.1mm$ waveguide with a peak pump power of less than $1W$
\cite{Cestier2012ol}. The present PSA uses a signal degenerate configuration
and exhibits a PSER of $10dB$ with total peak pump powers of less
than $0.5W$. The experimental results were confirmed by a simulation
using the split step Fourier transform (SSFT) numerical scheme.

\section{Results}

\begin{figure}[h]
\includegraphics[width=1\columnwidth]{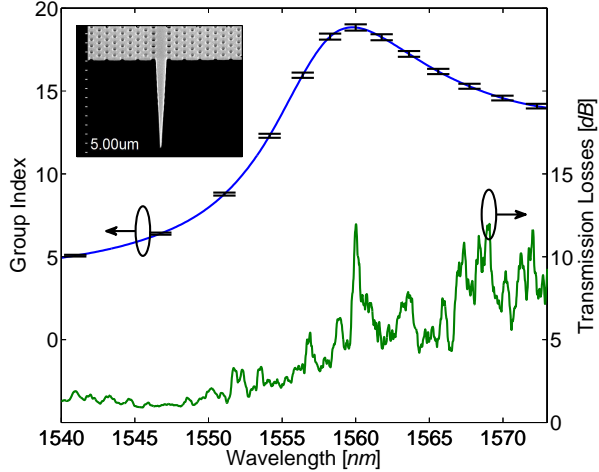}\caption{\label{fig:loss-ng}Spectra of group-index dispersion with estimated error-bars from calculations (blue curve and black bars,
left axis) and transmission losses (green curve, right axis) of the
dispersion engineered PCW. Inset: SEM image of the device input facet.}
\end{figure}

The device we tested consists of a $180nm$ thick $Ga_{0.5}In_{0.5}P$
membrane with air holes patterned in a hexagonal grid. Dispersion engineering \cite{Colman2012opex}
is obtained by shifting asymmetrically the two innermost rows next
to the core, and also changing the radius of the holes closest to
the core. Tapered mode converters are formed at the input and output
facets to increase the coupling efficiency and to prevent reflections. The inset of Fig. \ref{fig:loss-ng} is an SEM image of one waveguide end. The losses were
determined from transmission measurements of a wideband light source;
these are presented in Fig. \ref{fig:loss-ng}. The irregular shape of the transmission function stems from a cascade of micro defects that act as local resonances \cite{Mann2013ol}.
The group-index values were measured with accuracy of $1\%$ using the
OCT technique \cite{Caer2014apl}, and recalculated with the PCW
nonlinear properties using a periodic Finite Differences
Time Domain, with an accuracy of $0.1\%$. The errors are marked in Fig. 1 in black bars, from which a continuous function (blue curve) was evaluated using spline interpolation.

\begin{figure}[h]
\includegraphics[width=1\columnwidth]{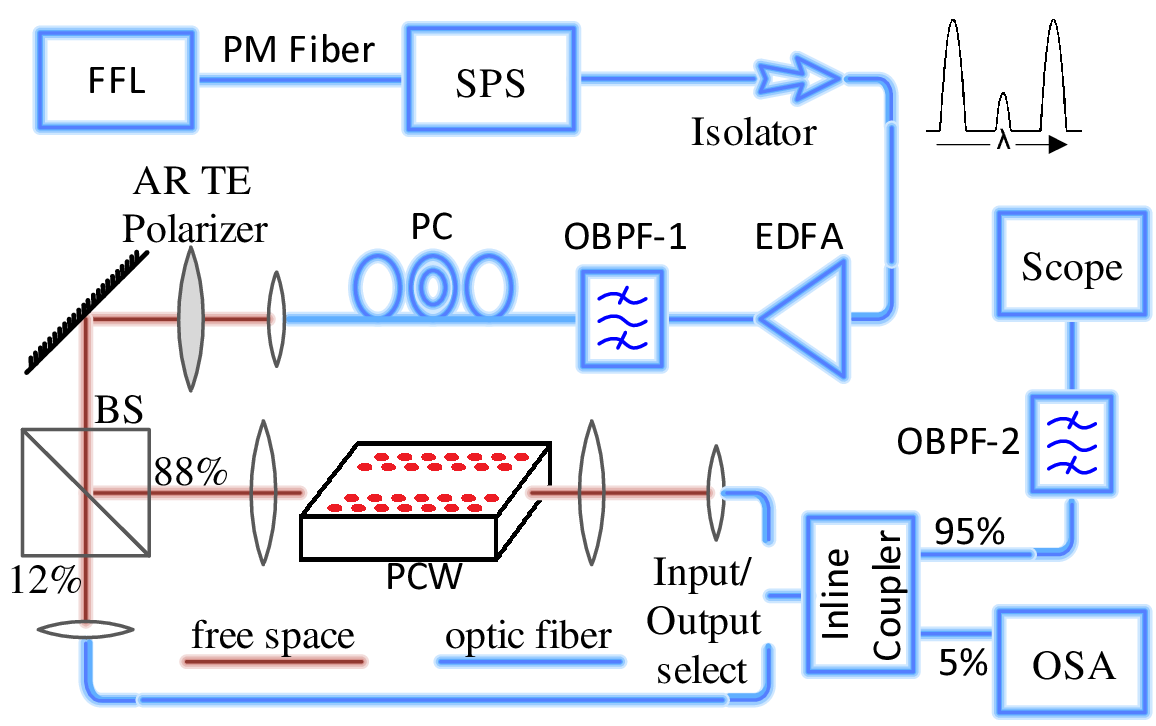}

\caption{\label{fig:Experimental-Setup}Experimental Setup.}
\end{figure}

The experimental setup is described in Fig. \ref{fig:Experimental-Setup}.
A Pritel femtosecond fiber laser (FFL) produces $1.4ps$ wide pulses at
a repetition rate of $R=20MHz$. The pulse's spectrum is sliced by
a Spectral Pulse Shaper (SPS) forming three coherent pulses at different
wavelengths. The amplitude and phase of each pulse is controllable
so that a signal-degenerate spectrum can be formed. Each pulse is
approximately $60ps$ wide with the signals at the longest and shortest
wavelengths serving as pumps. The experiments are performed with pulsed pumps since the membrane type PCWs cannot handle large average powers. Continuous wave (CW) operation is feasible by improving thermal conductivity or coating the PCW's surface by a dielectric to reduce surface effects.

The three waves are amplified by an Erbium-Doped Fiber Amplifier (EDFA).
The EDFA is followed by an Optical Bandpass Filter, designated OPBF-1,
that suppresses any FWM products that are generated within the EDFA.
The three waves propagate in free-space passing through an anti-reflection
(AR) polarizer. A beam-splitter (BS) extracts $12\%$ of the beam for monitoring
while the remaining $88\%$ is coupled to the PCW using an objective
lens. Outcoupling uses a similar lens after which the beam is coupled
to a single mode fiber which feeds the measurement instruments. An
optical spectrum analyzer (OSA) is used to examine the complete optical
spectrum from which the signal pulse is extracted by a narrowband
optical filter, designated OBPF-2 so it can be measured in the time
domain by a fast detector and a sampling oscilloscope.

\begin{figure}[h]
\includegraphics[width=1\columnwidth]{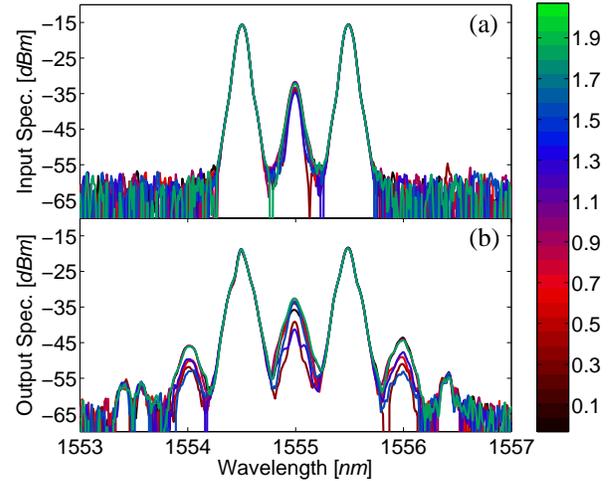}\llap{\parbox[b]{1in}{(a)\\\rule{0ex}{2.37in}}}\llap{\parbox[b]{1in}{(b)\\\rule{0ex}{1.25in}}}
\caption{\label{fig:spectra}Measured input (a) and output (b) spectra. The color coding represents
signal input phase in units of $\pi$ rad.}
\end{figure}

In the first set of measurements we describe, the three fields propagated
in the anomalous dispersion region of the waveguide. The signal wavelength
was $1555nm$ and the two pumps were detuned from it by $0.5nm$.
The EDFA operated at a gain level of $15dB$ which yielded a peak
power for each pump of $22dBm$, namely the total peak pump power was
$0.316W$. The SPS changed the signal phase from $0$ to $2\pi$ rad
in increments of $0.1\pi$ rad. Exemplary spectra are presented in Fig.
\ref{fig:spectra}(a)-\ref{fig:spectra}(b) for different signal
phase values. It is clear that FWM takes place in the $1.5mm$ long
semiconductor PCW with first and second order FWM products clearly
seen in Fig. \ref{fig:spectra}(b). Some degree of phase sensitive FWM takes place in the EDFA affecting the signal pulse shape and spectrum at the PCW input. We estimate that the pump peak powers are $16dB$ to $19dB$ stronger than that of signal at the PCW input.

\begin{figure}[h]
\includegraphics[width=1\columnwidth]{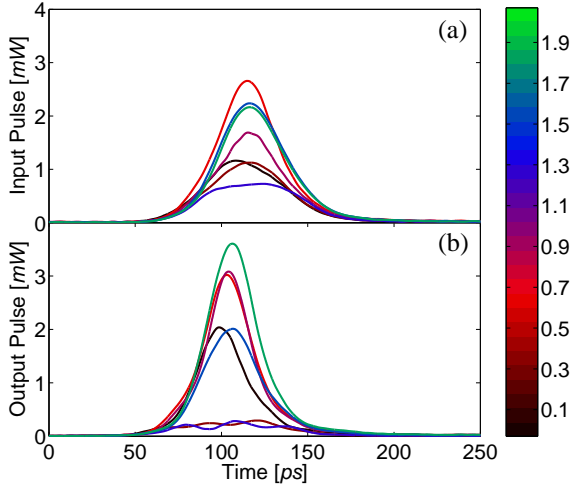}\llap{\parbox[b]{1in}{(a)\\\rule{0ex}{2.37in}}}\llap{\parbox[b]{1in}{(b)\\\rule{0ex}{1.25in}}}
\caption{\label{fig:pulses}Time domain measurement of input (a) and output (b) signal pulse envelopes.
The color coding represents signal input phase in units of $\pi$ rad.}
\end{figure}

Time domain measurements of the signal pulses at the input and output of the PCW are shown in Fig. \ref{fig:pulses}(a)-\ref{fig:pulses}(b). The $\pm4.2ps$ walk-off between the signal and pumps is small compared to the $60ps$ width of each pulse and has little effect on signal distortion.
The change in output signal pulse shape is due to the parametric gain dependence on the pump pulsed profile: (de-)amplification of the pulse center more than the tails causes (broadening) compression of the signal profile. Using CW or quasi-CW pumps can diminish signal distortions and improve bandwidth limitation.

\begin{figure}[h]
\includegraphics[width=1\columnwidth]{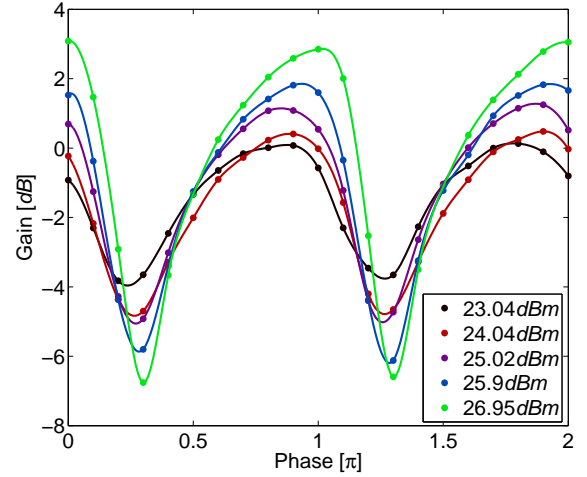}
\caption{\label{fig:Measured-gain}Measured phase-dependent net gain (dots) with
interpolated curves, for different total peak pump powers. The purple curve describes the gain matching the spectra and envelopes in Fig. \ref{fig:spectra} and Fig. \ref{fig:pulses} respectfully.}
\end{figure}

We thus examine the mean power gain given by
$G\left(\theta\right)=\overline{P}_{out}\left(\theta\right)/\overline{P}_{in}\left(\theta\right)$
with the mean power of the signal pulses defined
as $\overline{P}=R\int P(t)dt$. The integration is performed
over an oscilloscope span of less than $0.5ns$ (to account for the low duty-cycle
of the pulse train). The PCW input and output coupling
losses were determined to be $3.6dB$ and $1dB$, respectively. These
loss values are consistent with results of detailed simulations described
later in this paper. The phase-dependent net gain (inside the PCW) is presented in Fig. \ref{fig:Measured-gain}.
The green curve corresponds to the highest total peak pump powers
we used, $0.5W$. The other curves represent lower powers obtained
by reducing the EDFA gain. The gain varies cyclically with a period
of $\pi$ rad, and the PSER increases from $4dB$ to $10dB$ as the
power of the pumps increases. We note that for large peak pump powers, the minimum signal gain may be lower than shown in the figure since the lowest measurable signal level was limited by the receiver noise.

\begin{figure}[h]
\includegraphics[width=1\columnwidth]{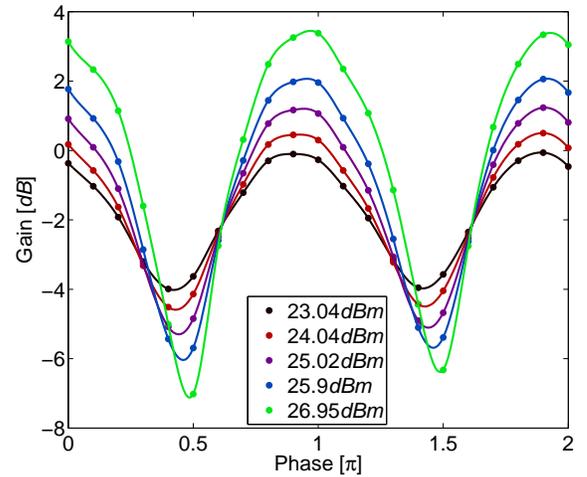}
\caption{\label{fig:Calculated-gain}Calculated phase-dependent net gain (dots)
with interpolated curves, for different total peak pump powers.}
\end{figure}

Pulse evolution upon propagation along the PCW was simulated using
the SSFT computation method with an adaptive step-size \cite{Menyuk2003jlt}.
The measured dispersion profile and loss spectrum were used in the
simulations together with the measured input pulse envelopes and spectra.
The nonlinear parameter at $1555nm$ was evaluated as $\gamma=1.5\,W^{-1}/mm$
using the model detailed in \cite{Roy201202PhtJrnl}. The small, $0.5nm$,
detuning between pumps and signal changes $\gamma$ insignificantly
so that the standard SSFT method suffices and a more accurate but
cumbersome technique \cite{Willinger2012JLT} is not needed. Different
initial signal phase values yield a calculated cyclical gain from
which the maximum gain and PSER were computed. An optimization procedure
for the input power of the waves and the exact loss level (which has
many random spectral features) yielded the best fit to experiments
when the input and output coupling losses were $3.6dB$ and $1dB$,
respectively. Figure \ref{fig:Calculated-gain} shows the calculated
signal gain dependence on phase for different total peak pump powers
which fit well with the measured results shown in Fig. \ref{fig:Measured-gain}.

The curves in Fig. \ref{fig:Measured-gain} show an asymmetry and differ from the theoretical logarithmic cosine dependence of the gain on the signal phase \cite{McKinstrie2006optcom}. Indeed the latter is deduced for a theoretical scenario assuming undepleted pumps and a weak signal, a condition not fully satisfied here. Moreover, prior to entering the PCW the three emitted waves propagate in an EDFA and through additional dispersive components so the input phases of the two pumps may not be the same. Finally, as can be seen in Fig. \ref{fig:spectra}(b), the additional idler waves also contribute to the asymmetry and reduce the the overall interaction between the two pumps and the signal \cite{Gao2012ol}. The simulations show a slight asymmetry, yet some of the parasitic effects are not considered and reconstruction of the experimental results is imperfect.

\begin{figure}[h]
\includegraphics[width=1\columnwidth]{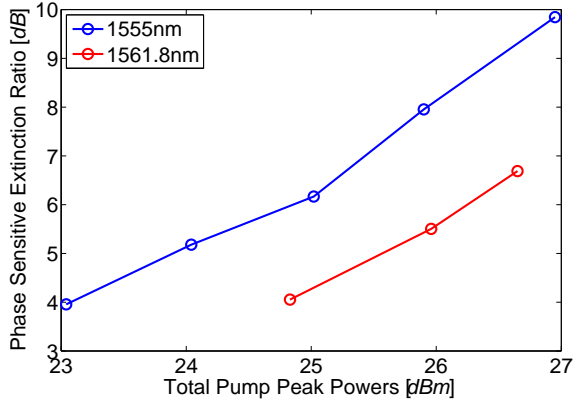}

\caption{\label{fig:Gain-Extinction-Ratio}PSER for different total peak pump
powers, for different signal wavelengths.}
\end{figure}

The measurements were repeated for a signal wavelengths of $1561.8nm$
with both pump detunings set once more to $0.5nm$. The waves propagate
in this case in the normal dispersion regime, where both the inherent
losses and the nonlinearity are larger ($\gamma=1.98\,W^{-1}/mm$).
The corresponding simulations show a similarly good fit to the experiments
when using the same coupling losses. The dependence of PSER on total
peak pump powers is shown in Fig. \ref{fig:Gain-Extinction-Ratio}.
The increased losses dominate the behavior at long wavelengths and
even though the nonlinearity increases, the overall phase-sensitive
interaction in the PCW decreases.

\section{Conclusions}
We have demonstrated phase sensitive parametric interactions
in a chip scale $1.5mm$ long $Ga_{0.5}In_{0.5}P$ dispersion engineered
PCW. A signal-degenerate configuration with two equally detuned pumps
produces PSER of $10dB$ with total pump peak powers of only $0.5W$. The moderate required pump power
is strictly due to the wide bandgap of $Ga_{0.5}In_{0.5}P$ which prevents TPA.
Higher PSER may be obtained with higher pump powers, however, membrane
structures tend to damage at high power and therefore the peak pump
powers in the present experiments were limited. Moreover, even though the absolute maximum net gain is moderate, this experiment proves that
mode squeezing takes place which can be used for phase regeneration
of noisy signals in phase encoded coherent communication.

\bibliographystyle{IEEEtran}
%\bibliography{references}
% Generated by IEEEtran.bst, version: 1.14 (2015/08/26)

\end{document}